\documentclass[reprint,superscriptaddress,aps,prl]{revtex4-2}
\usepackage{hyperref}
\usepackage{dcolumn}
\usepackage{bm}
\usepackage{braket}
\usepackage{physics}
\usepackage{subfigure}
\usepackage{graphicx}
\usepackage{epsfig}
\usepackage{amsmath, amssymb}
\usepackage{mathrsfs}

\usepackage{multirow}
\newcommand{\nn}{\nonumber}

\hypersetup{
setpagesize=false,
 bookmarksnumbered=true,%
 bookmarksopen=true,%
 colorlinks=true,%
 linkcolor=black,
 citecolor=blue,
}

\begin{document}
\preprint{APS/123-QED}
\title{
Proximity-Induced Giant Majorana Spin Response in a UTe$_2$--Metal Film Heterostructure
}

\author{Jushin Tei}
\email{tei@blade.mp.es.osaka-u.ac.jp}
\affiliation{Department of Materials Engineering Science, The University of Osaka, Toyonaka, Osaka 560-8531, Japan}
\author{Makoto Shimizu}
\affiliation{Department of Materials Engineering Science, The University of Osaka, Toyonaka, Osaka 560-8531, Japan}
\author{Yukio Tanaka}
\affiliation{Department of Applied Physics, Nagoya University, Nagoya 464-8603, Japan}
\author{Satoshi Fujimoto}
\affiliation{Department of Materials Engineering Science, The University of Osaka, Toyonaka, Osaka 560-8531, Japan}
\affiliation{Center for Quantum Information and Quantum Biology, The University of Osaka}
\affiliation{Center for Spintronics Research Network, Graduate School of Engineering Science, The University of Osaka}
\author{Takeshi Mizushima}
\email{mizushima@omu.ac.jp}
\affiliation{Department of Physics, Osaka Metropolitan University, Osaka 558-8585, Japan}
\affiliation{Nambu Yoichiro Institute of Theoretical and Experimental Physics (NITEP), Osaka Metropolitan University, Sumiyoshi, Osaka 558-8585, Japan}
\affiliation{Department of Materials Engineering Science, The University of Osaka, Toyonaka, Osaka 560-8531, Japan}
\date{\today}

\begin{abstract}
We establish an experimentally accessible route to identifying Majorana surface states through their intrinsic Ising spin response, using UTe$_2$ as a concrete platform.
The key is to extend the Majorana correlations into an adjacent metallic film via the superconducting proximity effect, thereby making the otherwise surface-localized response accessible to NMR measurements.
Remarkably, disorder further sharpens the Majorana signature:
impurity scattering suppresses the conventional $p$-wave proximity correlations while preserving the Majorana contribution, yielding a giant Ising-anisotropic enhancement of the NMR relaxation rate.
This approach also offers a powerful means to distinguish among candidate pairing symmetries.

\end{abstract}

\maketitle

\begin{table}[t]
\caption{Majorana surface states, their protecting crystalline symmetries, and the corresponding Majorana Ising-spin axes on the UTe$_2$$(100)$ surface.}
\centering
\begin{tabular}{lcccc}
\hline\hline
                        & $A_u$         & $B_{1u}$       & $B_{2u}$       & $B_{3u}$  \\ \hline
Majorana surface state & Dirac cone & Fermi arc & Fermi arc & None \\
Crystalline symmetry & $\mathcal{C}_a$ & $\mathcal{M}_{ca}$ & $\mathcal{M}_{ab}$ & --- \\
Ising-spin axis    & ${a}$          & ${b}$         & ${c}$         & ---    \\ \hline\hline
\end{tabular}
\label{table1}
\end{table}

\begin{figure*}[t]
    \centering
    \includegraphics[width=\linewidth]{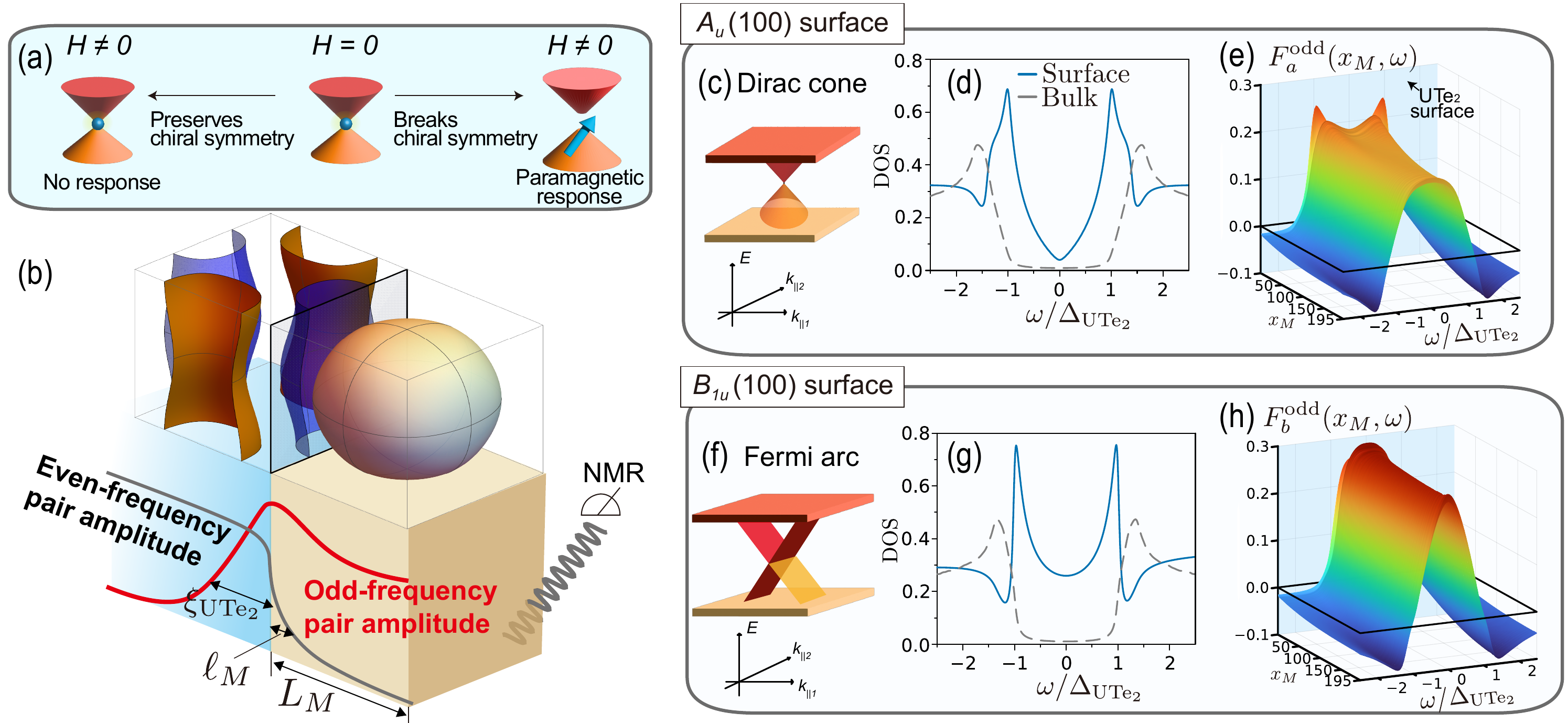} 
    \caption{(a)~Concept of Majorana Ising spin response.
    (b)~Schematic image of the UTe$_2$$(100)$--normal-metal film heterostructure.
    Odd-frequency pairing correlations associated with the Majorana surface states penetrate the metal film through the superconducting proximity effect and remain robust against nonmagnetic impurity scattering.
    NMR measurements on the metal film probe the proximity-enhanced Majorana Ising spin response.
    (c)~Majorana Dirac cone for the $A_u$ state on the $(100)$ surface.
    (d)~Corresponding surface DOS.
    (e)~Spatial and frequency dependence of the proximity-induced odd-frequency pair amplitude $F_a^{\rm odd}$ in the metal film coupled to UTe$_2$ in the $A_u$ state.
    (f)~Majorana Fermi arc for the $B_{1u}$ state on the $(100)$ surface.
    (g)~Corresponding surface DOS.
    (h)~Spatial and frequency dependence of the proximity-induced odd-frequency pair amplitude $F_b^{\rm odd}$ in the metal film coupled to UTe$_2$ in the $B_{1u}$ state.
    The results in (e) and (h) are calculated for $N_M=200$ and $\ell_M/a=10$.
    $x_M$ denotes the lattice coordinate in the metal film.}
    \label{fig1}
\end{figure*}

Over the past two decades, extensive efforts have sought to establish topological superconductivity and detect the associated Majorana quasiparticles~\cite{Read2000paird,kitaev2001unpaired,Fu2008superconducting,sato2009topological,Sato2009non,sato2010topological,Fu2010odd,Lutchyn2010Majorana,Oreg2010Helical,sato2011topology,alicea2012new,Mourik2012Signatures,das2012zero,deng2012anomalous,Stanescu2013Majorana,sato2016majorana}.
Nevertheless, a central challenge in candidate bulk topological superconductors is how to establish their topological nature experimentally.
Spectroscopic signatures of low-energy states alone do not necessarily provide conclusive evidence for their Majorana origin, motivating probes of intrinsic Majorana properties beyond the excitation spectrum~\cite{Sasaki2011Topological,Levy2013Experimental,Du2015Fully,ge2023single,prada2020andreev}.
One such property is the Ising-anisotropic Majorana spin response, whose experimental detection has remained challenging~\cite{Chung2009Detecting,Nagato2009Strong,vol10,Shindou2010Quantum,
Mizushima2012Symmetry}.
Here, we propose an experimentally feasible topological superconductor--metal film heterostructure that amplifies this response and enables its detection by nuclear magnetic resonance (NMR).

We focus on UTe$_2$, a leading candidate for spin-triplet and topological superconductivity~\cite{ran2019nearly,Aoki2019unconventional,aoki2022unconventional}.
Yet, its superconducting pairing symmetry remains unresolved despite extensive experimental efforts~\cite{Metz2019point,ishihara2023chiral,Iguchi2023Microscopic,suetsugu2024fully,theuss2024single,lee2025anisotropic,hayes2025robust,hayes2025field,zixuan2025obserbation,gu2025pair,wang2025imaging,Totsuka2026nodal,hosoi2026pseudo}.
Nevertheless, theoretical studies have shown that all possible pairing symmetries can support topological crystalline superconductivity and associated Majorana surface states (MSSs)~\cite{Tei2023Possible,tei2025topological}.
These MSSs are protected by chiral symmetries that combine the intrinsic Bogoliubov--de Gennes (BdG) chiral symmetry with either a twofold-rotation or a mirror symmetry.
Their existence and dispersion therefore depend sensitively on both the pairing symmetry and the surface orientation.

These chiral symmetries constrain the low-energy spin response of the MSSs to an Ising form~\cite{Mizushima2012Symmetry,shiozaki2014topology,volovik2010topological,Ueno2013Symmetry,Tsutsumi2013UPt3,xiong2017anisotropic,ohashi2024anisotropic}.
Specifically, the MSSs couple at low energies only to the magnetic-field component that breaks the protecting chiral symmetry~[Fig.~\ref{fig1}(a)].
Accordingly, the Majorana Ising axis in UTe$_2$ is fixed by the crystalline symmetry protecting the MSSs, as summarized for the $(100)$ surface in Table~\ref{table1}.
Although several schemes have been proposed to detect this characteristic response~\cite{Chung2009Detecting,Nagato2009Strong,Shindou2010Quantum,Higashitani2013Magnetic}, its direct observation remains challenging because the MSSs are confined to a thin surface region of order the superconducting coherence length, $\xi_{\rm UTe_2}\sim10~\mathrm{nm}$ in UTe$_2$~\cite{Rosuel2023Field}.
Consequently, the MSS contribution is too small to be resolved in conventional NMR measurements.

To overcome this challenge, we propose a UTe$_2$--normal-metal film heterostructure~[Fig.~\ref{fig1}(b)] and demonstrate that NMR measurements on the metal film can detect a strongly enhanced MSS-induced spin response.
In this heterostructure,
the superconducting proximity effect transfers the anomalous pair correlations associated with the MSSs into the film~\cite{DE1964Boundary,pannetier2000andreev,belzig1999quasiclassical}.
This leakage is encoded in proximity-induced odd-frequency pairing~\cite{berezinskii1974new,Tanaka2005Anomalous,Tanaka2007Theory,Linder2019Odd}.
Chiral-symmetry-protected MSSs are accompanied by a characteristic low-frequency odd-frequency component of the anomalous Green's function whose spectral structure is tied to that of the MSSs~\cite{Tamura2019Odd,Daido2019Chirality,tanaka2021theory}.
The induced odd-frequency pairs inherit the symmetry-constrained spin structure of the parent MSSs, thereby generating an Ising-anisotropic Majorana spin response throughout the metal film.
Owing to the large Fermi velocity of conventional metals, the proximity-induced correlations can extend over length scales exceeding NMR-experimentally accessible film thicknesses of $L_M=50$--$200~\mathrm{nm}$~\cite{Alfonsov2015Alfonsov,Lou2023Suppression,Beaudoin2025Superconducting}.
Moreover, the odd-frequency spin-triplet correlations contain an even-parity $s$-wave component that survives nonmagnetic impurity scattering, whereas the anisotropic even-frequency $p$-wave components are suppressed by impurity averaging~\cite{Tanaka2004Anomalous,Tanaka2005Anomalous,Bergeret2005Odd,Tanaka2007Theory,higashitani2009proximity,Tanaka2012Symmetry}.
Consequently, the odd-frequency $s$-wave component dominates in the diffusive regime, producing a pronounced and spatially extended Majorana Ising response.

We calculate the spatially resolved spin susceptibility $\chi^\alpha(x_M,x_M';\bm{q},\Omega)$ for a tight-binding model of the UTe$_2$--normal-metal film heterostructure and evaluate two NMR observables: the spin-lattice relaxation rate $1/T_1T$ and the Knight shift.
We find that the proximity-induced Majorana correlations produce characteristic Ising-anisotropic signatures in both observables. 
In particular, $1/T_1T$ is further enhanced in the diffusive regime, resulting in a giant Majorana-induced NMR signal.
The observation of this symmetry-locked Ising response provides compelling evidence for MSSs and could help distinguish among the candidate odd-parity pairing symmetries of UTe$_2$.

\textbf{\textit{Proximity-induced odd-frequency pairing.}}---
We consider a metal slab coupled to the $(100)$ surface of semi-infinite UTe$_2$.
Results for other high-symmetry surfaces are presented in the Supplemental Material~\cite{supply}.
The Green's function of the metal slab is given by
\begin{align}
    &\hat{G}_{x_Mx_M'}^{-1}(\bm{k}_{\parallel},\omega) = \omega\delta_{x_M,x_M'} - \hat{H}_{x_Mx_M'}(\bm{k}_{\parallel}) \nn \\
    &\hspace{10pt} - \delta_{x_M,x_M'}\hat{\Sigma}_{\rm imp}(x_M,\omega) -\delta_{x_M,1}\delta_{x_M',1}\hat{\Sigma}_{\rm UTe_2}(\bm{k}_{\parallel},\omega),
    \label{eq:Dyson}
\end{align}
where $\omega\in\mathbb{C}$ is a complex 
frequency, $\bm{k}_{\parallel}=(k_b,k_c)$ is the momentum parallel to the interface, and $x_M=1,\ldots,N_M$ labels the metal layers, with $x_M=1$ denoting the layer adjacent to UTe$_2$.
The number of metal layers is given by $N_M=L_M/a$, where $a$ is the lattice constant.
Here, $H_{x_Mx_M'}$ is the BdG Hamiltonian of the isolated metal slab, whose spherical Fermi surface is shown in Fig.~\ref{fig1}(b).
The impurity self-energy $\hat{\Sigma}_{\rm imp}$ is evaluated within the self-consistent Born approximation:
$\hat{\Sigma}_{\rm imp}(x_M,\omega) = \Gamma_{\rm imp}\sum_{\bm{k}_{\parallel}} \tau_z\hat{G}_{x_Mx_M}(\bm{k}_{\parallel},\omega)\tau_z$,
where $\Gamma_{\rm imp}=n_{\rm imp}V_{\rm imp}^2$ and $\tau_z$ is a Pauli matrix in Nambu space.
We characterize the disorder strength by
$\gamma_{\rm imp}=\pi\Gamma_{\rm imp}\rho_M=\hbar/(2\tau_{\rm imp})$, or equivalently by the mean free path $\ell_M=v_M\tau_{\rm imp}$.
In the diffusive regime, $\gamma_{\rm imp} \gg \Delta_{\rm UTe_2}$ and $\xi_{\rm diff} \gg \ell_M  $, where $\Delta_{\rm UTe_2}$ denotes the superconducting gap of the parent UTe$_2$.

The interface self-energy induced by coupling the metal to the UTe$_2$ $(100)$ surface is given by
$\hat{\Sigma}_{\rm UTe_2}(\bm{k}_{\parallel},\omega)
= \mathcal{T}^\dagger
\hat{G}_{\rm UTe_2}(\bm{k}_{\parallel},\omega)
\mathcal{T}$,
where $\mathcal{T}$ denotes the effective hopping matrix across the interface,
and $\hat{G}_{\rm UTe_2}$ is the surface Green's function of the semi-infinite UTe$_2$ system~\cite{ohashi2024anisotropic}.
The dependence on $\mathcal{T}$ is examined in Appendix~B, where we show that the following results remain robust over a wide range of interface transparencies, including the tunneling regime.
We employ a minimal two-band model with quasi-2D electron- and hole-like Fermi surfaces whose cylindrical axes lie along $k_c$, as observed by quantum-oscillation experiments and illustrated in Fig.~\ref{fig1}(b)~\cite{Aoki2022first,eaton2024quasi}.
To phenomenologically incorporate the heavy-fermion mass enhancement while preserving the Fermi-surface geometry, we rescale the normal-state band dispersion of UTe$_2$ by a quasiparticle weight $Z=0.01$.
For an ideal translationally invariant interface, the in-plane momentum $\bm{k}_{\parallel}$ is conserved, so that the overlap of the surface-projected Fermi surfaces controls the available phase space for proximity coupling.
The large projected Fermi surfaces of UTe$_2$ provide substantial momentum-space overlap with the conventional-metal Fermi surface considered here.
We neglect lattice mismatch, interfacial reconstruction, and interfacial disorder, which can introduce additional in-plane momentum-transfer channels and thereby relax the strict momentum-matching condition at realistic interfaces.

We consider all four odd-parity irreducible representations (IRs) of the $D_{2h}$ point group: $A_u$, $B_{1u}$, $B_{2u}$, and $B_{3u}$.
The $A_u$ state supports a Majorana Dirac cone on the $(100)$ surface, yielding the V-shaped surface density of states (DOS) shown in Figs.~\ref{fig1}(c) and \ref{fig1}(d)~\cite{Tei2023Possible}.
By contrast, the $B_{1u}$ and $B_{2u}$ states support Majorana Fermi arcs, which form 1D zero-energy manifolds in the surface Brillouin zone and produce a finite zero-energy surface DOS [Figs.~\ref{fig1}(f) and \ref{fig1}(g)]~\cite{yang2014dirac,sch15}.
These MSSs exhibit an Ising-anisotropic spin response, with the corresponding Ising axes listed in Table~\ref{table1}.
The $B_{3u}$ state does not support MSSs and does not show Majorana Ising spin responses on the $(100)$ surface.

The anomalous Green's function associated with these MSSs contains an odd-frequency component, which penetrates the normal-metal film through the superconducting proximity effect~\cite{Tanaka2005Anomalous,Tanaka2007Theory}.
Importantly, the corresponding penetration length exceeds the metal-film thicknesses considered here.
In the clean limit, it is characterized by the ballistic thermal coherence length
$\xi_{\rm bal}=\hbar v_M/{2\pi k_{\rm B}T}$~\cite{eilenberger1968transformation,BARZYKIN1999797}.
Because conventional metals generally have much larger Fermi velocities than heavy-fermion superconductors, $\xi_{\rm bal}$ can greatly exceed both the MSS localization length $\xi_{\rm UTe_2}$ and the experimentally accessible film thicknesses $L_M$.
In the diffusive regime, the relevant penetration length is instead
$\xi_{\rm diff}=\sqrt{\hbar D/(2\pi k_{\rm B}T)}$,
where $D=v_{M}\ell_M/3$ is the diffusion constant.
This length likewise remains larger than $L_M$
over the parameter range considered below~\cite{Usadel1970Generalized,BARZYKIN1999797}.
By fermionic antisymmetry, the induced odd-frequency spin-triplet correlations have even spatial parity and therefore include an $s$-wave component~\cite{berezinskii1974new}.
In the diffusive regime, impurity averaging suppresses the anisotropic even-frequency $p$-wave components, whereas the odd-frequency $s$-wave component survives and becomes dominant~\cite{Bakurskiy2014Anomalous}.

\begin{figure}[t]
    \centering
    \includegraphics[width=\linewidth]{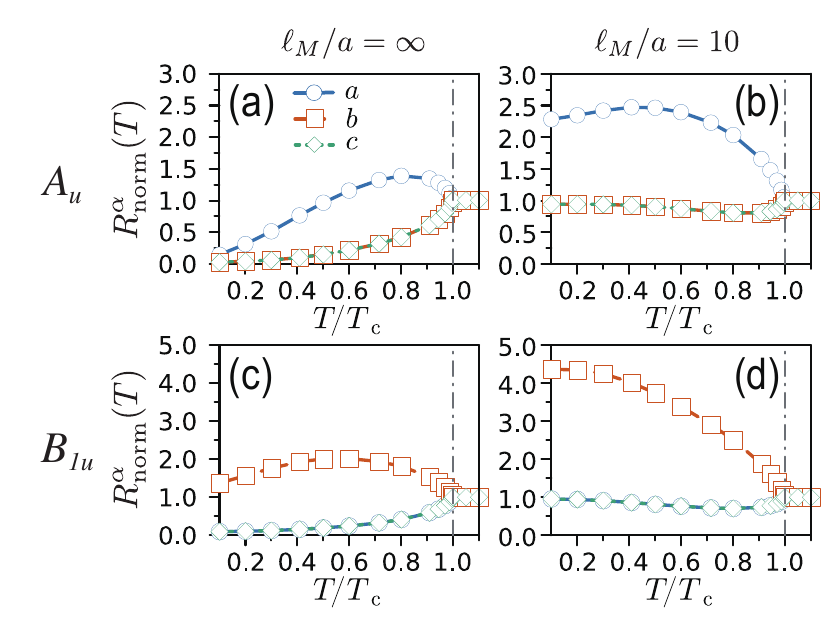} 
    \caption{
    Temperature dependence of the dynamical spin-response components $R^\alpha(T)$ for the $A_u$ [(a), (b)] and $B_{1u}$ [(c), (d)] states. Panels (a), (c) show the clean limit, and panels (b), (d) the diffusive regime, $\ell_M/a=10$.
    The metal thickness is fixed at $N_M=40$.
    The blue, orange, and green curves represent the response components along the ${a}$, ${b}$, and ${c}$ axes, respectively.
}
    \label{fig2}
\end{figure}

To demonstrate how these correlations penetrate the metal film, we define the local odd-frequency pair amplitude on the real-frequency axis as
\begin{align} 
    \hat{F}^{\rm odd}(x_M,\omega) = \frac{1}{2}\sum_{\bm{k}_\parallel} \left[ \hat{F}^R_{x_Mx_M}(\bm{k}_\parallel,\omega) - \hat{F}^A_{x_Mx_M}(\bm{k}_\parallel,-\omega) \right], 
\end{align}
where $\hat{F}^{R(A)}$ denotes the anomalous block of the retarded (advanced) Green's function obtained from Eq.~\eqref{eq:Dyson}.
The local odd-frequency component has spin-triplet symmetry and can be decomposed as
$\hat{F}^{\rm odd}=\bm{F}^{\rm odd}\cdot\boldsymbol{\sigma}i\sigma_b$,
where $\bm{F}^{\rm odd}=(F_a^{\rm odd},F_b^{\rm odd},F_c^{\rm odd})$ specifies its spin structure and $\sigma_{\alpha}$ $(\alpha = a, b, c)$ denotes a Pauli matrix in spin space.
At low frequencies, the MSS-induced odd-frequency pair amplitude has a nonzero component only along the Majorana Ising axis.

Figures~\ref{fig1}(e) and \ref{fig1}(h) show the calculated odd-frequency pair amplitudes induced in metal films coupled to UTe$_2$ in the $A_u$ and $B_{1u}$ states, respectively.
We consider a diffusive metal film with $\ell_M/a=10$ and $N_M=200$.
In the $A_u$ state, only $F_a^{\rm odd}$ is nonzero, consistent with the Majorana Ising axis along $a$.
Similarly, in the $B_{1u}$ state, only the Ising component $F_b^{\rm odd}$ is nonzero.
Remarkably, these odd-frequency amplitudes exhibit little spatial decay across the metal film, consistent with the proximity length exceeding the film thickness.
Moreover, the induced odd-frequency amplitude is larger in the $B_{1u}$ state than in the $A_u$ state, reflecting the greater low-energy spectral weight associated with the Majorana Fermi arc than with the Majorana Dirac cone.

\begin{figure}[t]
    \centering
    \includegraphics[width=\linewidth]{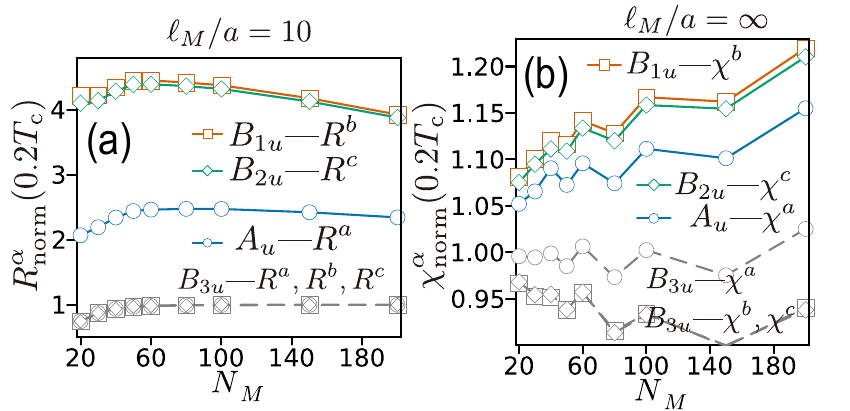} 
    \caption{(a)~Ising enhancement of the dynamical spin response $R^\alpha$ at $T=0.2T_{\rm c}$ in the diffusive regime with $\ell_M/a=10$.
    (b)~Ising enhancement of the uniform static spin susceptibility $\chi^\alpha$ at $T=0.2T_{\rm c}$ in the clean limit.
    In both panels, the metal thickness is varied from $N_M=20$ to $200$.
    Because the $B_{3u}$ state supports no Majorana surface states on the $(100)$ surface, none of the spin components exhibits an enhancement.}
    \label{fig3}
\end{figure}

\textbf{\textit{Giant Majorana Ising spin response in the relaxation rate.}}---
For a static magnetic field applied along the $\gamma$ axis, the spin-lattice relaxation rate is given by $1/T_1^\gamma T = A\lim_{\Omega\rightarrow 0}\sum_{x_M,\bm{q}_{\parallel}}\Im \chi^\pm_{\gamma}(x_M,x_M;\bm{q}_{\parallel},\Omega)/\Omega$,
where $A$ contains the hyperfine-coupling constant and the sum over $x_M$ accounts for the contribution from the entire metal film.
Because nuclear-spin relaxation is driven by spin fluctuations transverse to the applied field, we decompose the rate as
$1/T_1^\gamma T = R^\alpha(T) + R^\beta(T)$,
where $(\alpha,\beta,\gamma)$ is a permutation of $(a,b,c)$ and $R^\alpha(T)$ denotes the  dynamical spin response component along the $\alpha$ axis.
Its explicit form is given by
\begin{align}
    \label{eq:R}
    R^{\alpha} = \frac{A\hbar}{8T}\sum_{x_M}\int_{-\infty}^{\infty}\frac{d\omega}{2\pi}\cosh^{-2}\Big(\frac{\hbar\omega}{2k_{\rm B}T}\Big)
    \mathrm{tr}\Big[\hat{\mathcal{A}}_{x_M}\hat{S}^{\alpha}
    \hat{\mathcal{A}}_{x_M}\hat{S}^{\alpha}\Big],
\end{align}
where $\hat{\mathcal{A}}_{x_M}(\omega) = i\sum_{\bm{k}_{\parallel}}(\hat{G}^R-\hat{G}^A)_{x_Mx_M}$ and $\hat S^\alpha$ is the local spin operator.

Figures~\ref{fig2}(a) and \ref{fig2}(b) show the temperature dependence of $R^\alpha(T)$ for the $A_u$ state in clean and diffusive metal films, respectively. 
We define the normalized quantity as
$ R^\alpha_{\rm norm}(T) \equiv R^\alpha(T)/R^\alpha(T_{\rm c})$.
In the clean limit, the even-frequency contribution suppresses the dynamical spin response, whereas the MSS-induced odd-frequency component provides a paramagnetic contribution to the Majorana Ising component $R^a$. Consequently, $R^b$ and $R^c$ decrease monotonically upon cooling, while $R^a$ increases just below $T_{\rm c}$ before decreasing at lower temperatures. In the diffusive film, impurity scattering suppresses the anisotropic even-frequency contribution, leaving $R^b$ and $R^c$ nearly temperature independent, while the odd-frequency $s$-wave component survives and strongly enhances $R^a$, which exceeds $2R^a(T_{\rm c})$ at low temperatures.
The $B_{1u}$ state shows analogous behavior with the Majorana Ising axis along $b$ [Figs.~\ref{fig2}(c) and \ref{fig2}(d)]: the non-Ising components $R^a$ and $R^c$ are suppressed or weakly temperature dependent, whereas $R^b$ is strongly enhanced and reaches $4R^b(T_{\rm c})$. The larger enhancement than in the $A_u$ state reflects the greater low-energy spectral weight of the Majorana Fermi arc compared with the Majorana Dirac cone.
We note that the static Zeeman field is not explicitly included in the present linear-response calculation.
For the $1/T_1T$ measurement probing the Ising enhancement, the static field is perpendicular to the Majorana Ising axis, and we have numerically confirmed that its effect on the enhancement is negligible.

Figure~\ref{fig3}(a) summarizes the Majorana-induced enhancement of the Ising component $R^\alpha$ at $T=0.2T_{\rm c}$ for all IRs in the diffusive regime with $\ell_M/a=10$, as the metal thickness is varied from $N_M=20$ to $200$.
The enhancement is nearly independent of the metal thickness.
This robustness suggests that a sizable Majorana signal should remain observable for metal-film thicknesses suitable for NMR measurements.
For metal films coupled to the $(100)$ surface of UTe$_2$ in the $A_u$, $B_{1u}$, and $B_{2u}$ states, the response component along the corresponding Majorana Ising axis is strongly enhanced.
By contrast, in the $B_{3u}$ state, which supports no symmetry-protected MSSs on the $(100)$ surface, all three components remain close to their normal-state values.

\begin{figure}[t]
    \centering
    \includegraphics[width=\linewidth]{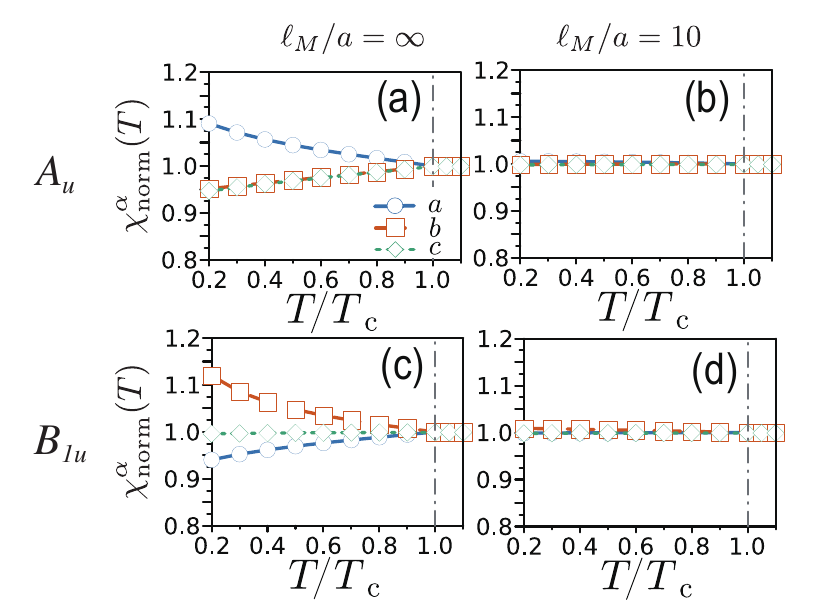} 
    \caption{
Temperature dependence of the uniform static spin susceptibility $\chi^\alpha(T)$ in the metal film for the $A_u$ [(a), (b)] and $B_{1u}$ [(c), (d)] states. Panels (a),(c) show the clean limit and panels (b), (d) the diffusive regime, $\ell_M/a=10$.
    }
    \label{fig4}
\end{figure}

\textbf{\textit{Majorana Ising spin response in the Knight shift.}}---
We next consider the static spin susceptibility of the metal film, which is directly relevant to Knight-shift measurements.
The uniform static susceptibility along the $\alpha$ axis is defined as
$\chi^\alpha(T)=\sum_{x_M,x_M'}\chi^\alpha(x_M,x_M';\bm{q}_{\parallel}=0,\Omega=0)$. 
Figures~\ref{fig4}(a) and \ref{fig4}(b) [Figs.~\ref{fig4}(c) and \ref{fig4}(d)] show its temperature dependence for the clean limit and the diffusive regime in metal films coupled to UTe$_2$ in the $A_u$ [$B_{1u}$] state.
A clear Ising-anisotropic response develops in the clean limit.
On the other hand, its temperature dependence is almost completely suppressed in the diffusive metal.
This behavior contrasts with the pronounced enhancement of the dynamical spin response discussed above.
To understand this contrasting behavior, we analytically evaluate both spin responses using a phenomenological model of the odd-frequency $s$-wave proximity effect (Appendix A).
We find that the enhancement of the static Ising susceptibility is suppressed at least as $\gamma_{\rm imp}^{-2}$.
In contrast, the dynamical spin response is highly sensitive to the low-energy Majorana spectral weight and can remain strongly enhanced even in the diffusive regime.

In the clean metal film, both even- and odd-frequency pairing correlations contribute to the static spin susceptibility.
For the $A_u$ state, $\chi^b$ and $\chi^c$ decrease upon cooling, reflecting the conventional response induced by spin-triplet pairing, whereas $\chi^a$ is enhanced by the paramagnetic contribution associated with the MSS-induced odd-frequency correlations.
A similar behavior is found in the $B_{1u}$ state: the non-Ising components, $\chi^a$ and $\chi^c$, exhibit temperature dependences characteristic of conventional spin-triplet pairing, whereas the Majorana Ising component $\chi^b$ is enhanced.
Figure~\ref{fig3}(b) summarizes the Majorana-induced enhancement of the Ising component of the static spin susceptibility at $T=0.2T_{\rm c}$ for all IRs in the clean limit.
Within the range of film thicknesses considered, the enhancement increases with increasing film thickness, reaching over $10\%$ for the $A_u$ state and over $15\%$ for the $B_{1u}$ and $B_{2u}$ states.
These sizable changes suggest that the Majorana Ising response may be detectable through Knight-shift measurements in sufficiently clean metal films.
Because the $B_{3u}$ state supports no MSSs on the $(100)$ surface, none of the spin-susceptibility components exhibits a corresponding enhancement.

\textbf{\textit{{Discussion}.}}---
In summary, we have demonstrated that superconducting proximity coupling provides a route to convert the intrinsically surface-localized Majorana spin response into a sizable NMR signal in an adjacent metal film.
The spin-lattice relaxation rate $1/T_1T$ exhibits a pronounced Majorana-induced Ising-anisotropic enhancement even in a diffusive metal film, whereas detection of the corresponding enhancement in the Knight shift requires a sufficiently clean metal film.

The proper choice of metal film materials is constrained by both efficient proximity coupling and experimental feasibility. 
A large Fermi velocity is desirable for extending the proximity-induced correlations throughout the film, while substantial overlap between the surface-projected Fermi surfaces of the metal and UTe$_2$ favors proximity coupling when in-plane momentum is approximately conserved. 
The metal should also remain normal over the relevant temperature and magnetic-field range, possess suitable NMR-active nuclei, and form a stable interface with UTe$_2$.

From the viewpoint of electronic structure, Li and Na are attractive because of their nearly spherical Fermi surfaces and large Fermi velocities of order $10^6\mathrm{\,m/s}$~\cite{gall2016electron}; NMR measurements are also well established~\cite{narath1968effects}. Their high chemical reactivity, however, may complicate interface fabrication. Al, Cu, and Ag are experimentally more practical alternatives: despite their more complex Fermi surfaces, they also have large Fermi velocities, established NMR protocols~\cite{knight1949nuclear,redfield1955nuclear,narath1966nuclear}, and mature thin-film fabrication. For Al, particular care is required because its superconducting transition temperature, approximately $1.2$ K in bulk, can be enhanced in thin films~\cite{cohen1968superconductivity,yeh2023doubling}; the film thickness and applied field must therefore be chosen to keep Al normal. Overall, Li and Na are electronically favorable, whereas Cu and Ag may be experimentally more practical, with Al also viable under appropriate conditions. The microscopic structure and chemical stability of the UTe$_2$--metal interface remain important experimental issues.

\textbf{\textit{Acknowledgments.}}---
The authors are grateful to K. Ishida and S. Kitagawa for valuable discussions from an experimental point of view.
J.T. is grateful to H. Kagawa and R. Miwa for fruitful discussions.
J.T. is supported by a Japan Society for the Promotion of Science (JSPS) Fellowship for Young Scientists.
This work was supported by JSPS KAKENHI (Grant No.~JP23K17668, No.~JP23K22492, No.~JP24KJ1621, No.~JP24K00556, No.~JP24K00578, No.~JP24K00583, No.~JP25H00599, No.~JP25H00609, No.~JP25K07227, and No.~JP25K22011) and a Grant-in-Aid for Transformative Research Areas (A) ``Correlation Design Science'' (Grant No.~JP25H01250) from JSPS of Japan.

\bibliography{paper}

\section*{End Matter}

\textbf{\textit{Appendix A: Analytical model for the spin response in the diffusive metal.}}---
We construct a phenomenological low-energy model for a diffusive metal attached to a topological superconductor.
In this regime, impurity scattering suppresses momentum-dependent pairing correlations, while the odd-frequency spin-triplet $s$-wave component associated with MSSs remains robust.
We therefore retain only this component in the proximity-induced self-energy.
Tamura \textit{et al.} established the spectral bulk-boundary correspondence, which relates the bulk winding number $W(\bm{k}_{\parallel},z)$ to the odd-frequency
pair amplitude at the boundary, $F^{\rm odd}(\bm{k}_{\parallel},z)\propto W(\bm{k}_{\parallel},z)/z$
~\cite{Tamura2019Odd,Mizushima2023odd}.
Furthermore, the self-conjugate nature of a Majorana zero mode requires equal normal and anomalous residues of the singular Green's function, up to a Nambu gauge transformation.
Motivated by these properties, we introduce a phenomenological self-energy for the metal,
\begin{align}
\hat\Sigma_{\rm UTe_2}(i\omega_n)
 =\frac{\lambda^2}{i\omega_n+i\kappa\,\mathrm{sgn}\,\omega_n}
   (\tau_0+\tau_x),
\label{eq:app_MSS_self_energy}
\end{align}
and the Green's function is determined by the Dyson equation
\begin{align}
    \hat G^{-1}(\bm{k},i\omega_n)
    =Z_n\tau_0-\xi_{\bm{k}}\tau_z-\hat\Sigma_{\rm UTe_2}(i\omega_n),
    \label{eq:app_Ginv}
\end{align}
where $Z_n=i[\omega_n+\gamma_{\rm imp}\,\mathrm{sgn}\,\omega_n]$,
$\xi_{\bm{k}}$ is the normal-metal dispersion measured from the Fermi level, $\bm{k}$ is the three-dimensional momentum defined by neglecting the presence of the interface, and $\tau_\mu$ are Pauli matrices in the reduced Nambu space.
Here, $\gamma_{\rm imp}$ is the impurity scattering rate in the metal, $\lambda$ characterizes the proximity coupling, and $\kappa$ represents the phenomenological linewidth of the MSS pole.
The linewidth $\kappa$ may originate from hybridization with continuum states or other relaxation mechanisms and is treated independently of $\gamma_{\rm imp}$.
The persistence of pronounced low-frequency odd-frequency pair amplitudes under strong impurity scattering, as shown in Figs.~\ref{fig1}(e) and \ref{fig1}(h), motivates considering a narrow MSS linewidth $\kappa\ll\gamma_{\rm imp}$.
Throughout this Appendix, we set $\hbar=k_{\rm B}=1$.

Defining
$C_n=\lambda^2/[i\omega_n+i\kappa\,\mathrm{sgn}\,\omega_n]$,
we obtain
\begin{equation}
\hat G(\bm{k},i\omega_n)=
\frac{(Z_n-C_n)\tau_0+\xi_{\bm{k}}\tau_z+C_n\tau_x}
{Z_n^2-\xi_{\bm{k}}^2-2Z_nC_n}.
\label{eq:app_G}
\end{equation}
We choose the spin basis such that the spin-diagonal anomalous component corresponds to the odd-frequency spin-triplet amplitude $F_y^{\rm odd}$, with $y$ denoting the Majorana Ising axis.
The corresponding spin vertices in the reduced Nambu space are $V_y=\tau_0$ and $V_x=V_z=\tau_z$.

Within the bubble approximation, the uniform static spin susceptibility is given by
\begin{equation}
\chi^\mu=-\frac{T}{4}\sum_{\omega_n,\mathbf k}
\mathrm{Tr}[V_\mu\hat G(\mathbf k,i\omega_n)
V_\mu\hat G(\mathbf k,i\omega_n)].
\label{eq:app_chi_def}
\end{equation}
In the wide-band limit with a constant normal-metal density of states $\rho_M$, the non-Ising components remain at their normal-state value, $\chi^x=\chi^z=\chi_N\equiv\rho_M/2$.
In contrast, the Ising component acquires an additional positive contribution, $\delta\chi^y\equiv\chi^y-\chi_N>0$.
Performing the integration over $\xi_{\bm{k}}$, we obtain
\begin{align}
\frac{\delta\chi^y}{\chi_N}
&=\pi T\lambda^4\sum_{\omega_n}
\frac{1}{(|\omega_n|+\kappa)^2D_n^{3/2}},
\label{eq:app_static_sum}\\
D_n&=(|\omega_n|+\gamma_{\rm imp})^2
+\frac{2\lambda^2(|\omega_n|+\gamma_{\rm imp})}
{|\omega_n|+\kappa}.
\end{align}
At $T=0$, in the strong-scattering regime
$\gamma_{\rm imp}\gg\lambda$ and
$\kappa\ll\gamma_{\rm imp}$, we obtain
\begin{align}
\frac{\delta\chi^y}{\chi_N}
\simeq
\frac{\lambda^2}{\gamma_{\rm imp}^2}
\left(
1-\sqrt{
\frac{\kappa}
{\kappa+2\lambda^2/\gamma_{\rm imp}}}
\right).
\label{eq:app_static_scaling}
\end{align}
For $\kappa\ll\lambda^2/\gamma_{\rm imp}$,
this reduces to
$\delta\chi^y/\chi_N\simeq
(\lambda/\gamma_{\rm imp})^2$.
Thus, the static Ising enhancement is strongly suppressed by impurity scattering, with a finite $\kappa$ leading to further suppression.

We next consider the local low-frequency dynamical spin response $R^\mu$.
Within the same bubble approximation, we have
\begin{equation}
R^\mu=\int d\omega\,[-f'(\omega)]
\mathrm{Tr}[V_\mu\hat A(\omega)V_\mu\hat A(\omega)],
\label{eq:app_R_def}
\end{equation}
where $\hat A(\omega)=-(1/\pi)\mathrm{Im}\,\sum_{\mathbf k}
\hat G^R(\mathbf k,\omega)$. 
With
$Z(\omega)=\omega+i\gamma_{\rm imp}$, $C(\omega)=\lambda^2/(\omega+i\kappa)$, and
$P(\omega)=\sqrt{Z^2-2ZC}$,
the local spectral function becomes
\begin{align}
\hat A(\omega)&=\rho_M[n(\omega)\tau_0+m(\omega)\tau_x],
\label{eq:app_A}\\
n(\omega)&=\mathrm{Re}\frac{Z(\omega)-C(\omega)}{P(\omega)},\qquad
m(\omega)=\mathrm{Re}\frac{C(\omega)}{P(\omega)}.
\end{align}
Consequently,
\begin{align}
R^{x}=R^z&=2\rho_M^2\int d\omega\,[-f'(\omega)](n^2(\omega)-m^2(\omega)),\\
R^y&=2\rho_M^2\int d\omega\,[-f'(\omega)](n^2(\omega)+m^2(\omega)).
\label{eq:app_R_nm}
\end{align}
At $T=0$, the response is determined by $\omega=0$. 
Defining the dimensionless parameter
$\zeta=\lambda^2/(\gamma_{\rm imp}\kappa)$,
we obtain
\begin{align}
n(0)&=\frac{1+\zeta}{\sqrt{1+2\zeta}},\\
m(0)&=-\frac{\zeta}{\sqrt{1+2\zeta}}.
\label{eq:app_nm_zero}
\end{align}
With $R_N=2\rho_M^2$, it follows that
$R^x=R^z=R_N$.
In contrast, the Ising component exhibits an enhancement,
\begin{equation}
\frac{\delta R^y}{R_N}
=\frac{2\zeta^2}{1+2\zeta},
\label{eq:app_R_enhancement}
\end{equation}
where $\delta R^y=R^y-R_N$.

In the asymptotic limit $\gamma_{\rm imp}\to\infty$ at fixed $\lambda$ and $\kappa$, the dynamical enhancement is suppressed as $\delta R^y/R_N\propto\gamma_{\rm imp}^{-2}$.
Nevertheless, a sufficiently small $\kappa$ allows a pronounced enhancement even in the strongly diffusive regime.
This contrast reflects the different sensitivities to the low-energy spectrum: the static susceptibility remains finite as $\kappa\to0$, whereas the dynamical response directly probes the zero-energy spectral weight at $T=0$, yielding an enhancement proportional to $1/\kappa$ for $\zeta\gg1$.
Thus, the spin-lattice relaxation rate provides a particularly sensitive probe of the low-energy Majorana Ising response in the diffusive regime.

\begin{figure}[t]
    \centering
    \includegraphics[width=\linewidth]{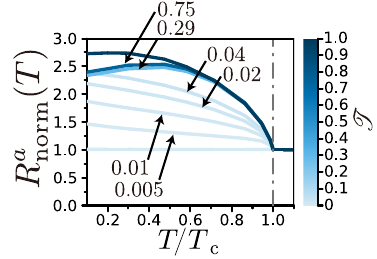} 
    \caption{
Temperature dependence of the normalized Ising spin response
$ R^a_{\rm norm}(T)$ for the $A_u$ state
for several effective interface transparencies $\mathscr{T}$.
A pronounced Ising enhancement persists even in the tunneling regime.
    }
    \label{figE5}
\end{figure}

\textbf{\textit{Appendix B: Robustness against an interface barrier.}---}
We examine the robustness of the Majorana spin response against a finite interface barrier.
To characterize the interface coupling, we introduce the effective transparency $\mathscr{T}\in[0,1]$ as
\begin{align}
    \label{eq:transparency}
    \mathscr{T}
    =
    \frac{4\pi^2|\mathcal{T}|^2\rho_{\rm UTe_2}\rho_M}
    {\left(1+\pi^2|\mathcal{T}|^2\rho_{\rm UTe_2}\rho_M\right)^2},
\end{align}
where $|\mathcal{T}|$ denotes the magnitude of the effective transfer integral across the interface.
Here, $\mathscr{T}=1$ corresponds to a perfectly transparent interface, while $\mathscr{T}\ll1$ represents the tunneling regime.

Figure~\ref{figE5} shows the temperature dependence of the normalized Ising component $ R^a_{\rm norm}(T)$ for the $A_u$ state at the (100) interface for several values of $\mathscr{T}$.
As expected, the Ising enhancement is gradually reduced as the interface transparency decreases.
Remarkably, however, a strong Majorana-induced spin response persists deep in the tunneling regime.
Even for $\mathscr{T}=0.01$, $R^a(T)$ reaches approximately twice its normal-state value at low temperatures.
Owing to the large separation between the electronic and superconducting energy scales, the effective transfer integral can remain sufficiently large compared with the superconducting gap even in the tunneling regime.
Consequently, a pronounced proximity-induced Ising response can survive even for an interface in the tunneling regime, substantially relaxing the requirement on interface transparency for the proposed NMR detection scheme.

\end{document}


\title{Supplemental Material for ``Proximity-Induced Giant Majorana Spin Response in a UTe$_2$--Metal Film Heterostructure''}
\author{Jushin Tei}
\email{tei@blade.mp.es.osaka-u.ac.jp}
\affiliation{Department of Materials Engineering Science, The University of Osaka, Toyonaka, Osaka 560-8531, Japan}
\author{Makoto Shimizu}
\affiliation{Department of Materials Engineering Science, The University of Osaka, Toyonaka, Osaka 560-8531, Japan}
\author{Yukio Tanaka}
\affiliation{Department of Applied Physics, Nagoya University, Nagoya 464-8603, Japan}
\author{Satoshi Fujimoto}
\affiliation{Department of Materials Engineering Science, The University of Osaka, Toyonaka, Osaka 560-8531, Japan}
\affiliation{Center for Quantum Information and Quantum Biology, The University of Osaka}
\affiliation{Center for Spintronics Research Network, Graduate School of Engineering Science, The University of Osaka}
\author{Takeshi Mizushima}
\email{mizushima@omu.ac.jp}
\affiliation{Department of Physics, Osaka Metropolitan University, Osaka 558-8585, Japan}
\affiliation{Nambu Yoichiro Institute of Theoretical and Experimental Physics (NITEP), Osaka Metropolitan University, Sumiyoshi, Osaka 558-8585, Japan}
\affiliation{Department of Materials Engineering Science, The University of Osaka, Toyonaka, Osaka 560-8531, Japan}
\date{\today}

\maketitle

\section{Tight-binding Hamiltonian}
Here, we describe the tight-binding models used for the UTe$_2$$(100)$--normal-metal film heterostructure.
The long-range proximity effect discussed in the main text is favored by two properties of the metal: a large Fermi velocity and substantial overlap with the surface-projected Fermi surfaces of UTe$_2$.
To model the strong mass renormalization in UTe$_2$, we introduce a phenomenological renormalization factor $Z$ that rescales its normal-state band dispersion.
For comparable bare velocity scales in the two materials, $Z$ approximately represents the characteristic Fermi-velocity ratio $v_{\rm UTe_2}/v_M$.

The normal-state dispersion of the metal is given by
\begin{align}
\xi_M(\bm{k})
={}&
8t_{m1}
\cos\frac{k_a}{2}
\cos\frac{k_b}{2}
\cos\frac{k_c}{2}
+
2t_{m2}
\left(
\cos k_a+\cos k_b+\cos k_c
\right)
-\mu_m .
\label{eq}
\end{align}
The large Fermi surface shown in Fig.~1(b) of the main text is obtained using
\begin{align}
t_{m1}=-1.0,\qquad
t_{m2}=-0.42,\qquad
\mu_m=-1.6.
\end{align}
All energies are measured in units of $|t_{m1}|$.

The normal-state Hamiltonian of UTe$_2$ is diagonal in the electron- and hole-band basis:
\begin{align}
\hat{\xi}_{\rm UTe_2}(\bm{k})
=
Z
\begin{pmatrix}
\xi_e(\bm{k})\otimes\sigma_0 & 0\\
0 & \xi_h(\bm{k})\otimes\sigma_0
\end{pmatrix},
\end{align}
where $\sigma_0$ is the identity matrix in spin space and
\begin{subequations}
\begin{align}
\xi_e(\bm{k})
={}&
2t_{e1}\cos k_a
+2t_{e2}\cos k_b
+2t_{e3}\cos k_c
+4t_{e4}\cos k_a\cos k_b
-\mu_e ,
\\
\xi_h(\bm{k})
={}&
2t_{h1}\cos k_a
+2t_{h2}\cos k_b
+2t_{h3}\cos k_c
+4t_{h4}\cos k_a\cos k_b
-\mu_h .
\end{align}
\end{subequations}
The factor $Z$ reduces the quasiparticle velocity without changing the Fermi-surface geometry.
We use $Z=0.01$.
The quasi-2D electron- and hole-like Fermi surfaces shown in Fig.~1(b) of the main text are obtained using
\begin{align}
&(t_{e1},t_{e2},t_{e3},t_{e4},\mu_e)
=
(-1.0,1.0,0.5,0.7,-2.0),\nn \\
&(t_{h1},t_{h2},t_{h3},t_{h4},\mu_h)
=
(1.0,-1.0,0.5,-0.7,2.0).
\end{align}

For simplicity, we neglect interband pairing and assume the same intraband gap function on the electron and hole bands.
The spin-triplet pairing matrix is
\begin{align}
\hat{\Delta}_{\lambda}(\bm{k})
=
\left[
\bm{d}(\bm{k})\cdot\boldsymbol{\sigma}
\right]i\sigma_b,
\qquad \lambda=e,h ,
\end{align}
where the $d$ vectors for the four odd-parity irreducible representations (IRs) are
\begin{align}
\bm{d}_{A_u}(\bm{k})
&=
\Delta_{\rm UTe_2}
\begin{pmatrix}
\sin k_a\\
\sin k_b\\
\sin k_c
\end{pmatrix},
&
\bm{d}_{B_{1u}}(\bm{k})
=
\Delta_{\rm UTe_2}
\begin{pmatrix}
\sin k_b\\
\sin k_a\\
0
\end{pmatrix},
\nonumber\\
\bm{d}_{B_{2u}}(\bm{k})
&=
\Delta_{\rm UTe_2}
\begin{pmatrix}
\sin k_c\\
0\\
\sin k_a
\end{pmatrix},
&
\bm{d}_{B_{3u}}(\bm{k})
=
\Delta_{\rm UTe_2}
\begin{pmatrix}
0\\
\sin k_c\\
\sin k_b
\end{pmatrix}.
\end{align}
Here, $\Delta_{\rm UTe_2}$ denotes the overall pairing amplitude,
which we set to $\Delta_{\rm UTe_2}=0.1Z$.
Its temperature dependence follows the BCS interpolation formula
$\Delta_{\rm UTe_2}(T)=\Delta_{\rm UTe_2}
\tanh[1.74\sqrt{T_{\rm c}/T-1}]$,
where $T_{\rm c}=\Delta_{\rm UTe_2}/(1.764k_{\rm B})$.
For simplicity, we assume that electrons in the metal couple with equal amplitudes to the electron- and hole-like bands of UTe$_2$.
The interfacial hopping matrix is written as
\begin{align}
\mathcal{T}
=
h
\begin{pmatrix}
1\\
1
\end{pmatrix}_{\rm band}
\otimes\sigma_0\otimes\tau_z,
\end{align}
where $h$ is the hopping amplitude across the interface, $\sigma_0$ is the identity matrix in spin space, and $\tau_z$ is a Pauli matrix in Nambu space.
We use $h=0.2$ in the main calculations.
The two entries of the band-space vector correspond to the electron- and hole-like bands of UTe$_2$, respectively.

\begin{table}[t]
  \caption{Summary of Majorana surface states, protecting crystalline operators, and Ising spin directions.}
  \label{table:UTe2_topo}
  \centering
  \begin{ruledtabular}
  \begin{tabular}{llccc}
    IR & Quantity & (100) & (010) & (001)   \\\hline
    $A_u$
      & Majorana surface states
      & Dirac cone
      & Dirac cone
      & --- \\
      & Crystalline symmetry
      & $\mathcal{C}_a$
      & $\mathcal{C}_b$
      & --- \\
      & Ising direction
      & $a$
      & $b$
      & --- \\\hline
    $B_{1u}$
      & Majorana surface states
      & Fermi arc
      & Fermi arc
      & --- \\
      & Crystalline symmetry
      & $\mathcal{M}_{ca}$
      & $\mathcal{M}_{bc}$
      & --- \\
      & Ising direction
      & $b$
      & $a$
      & --- \\\hline
    $B_{2u}$
      & Majorana surface states
      & Fermi arc
      & ---
      & Fermi arc \\
      & Crystalline symmetry
      & $\mathcal{M}_{ab}$
      & ---
      & $\mathcal{M}_{bc}$ \\
      & Ising direction
      & $c$
      & ---
      & $a$ \\\hline
    $B_{3u}$
      & Majorana surface states
      & ---
      & Fermi arc
      & Fermi arc \\
      & Crystalline symmetry
      & ---
      & $\mathcal{M}_{ab}$
      & $\mathcal{M}_{ca}$ \\
      & Ising direction
      & ---
      & $c$
      & $b$ \\
  \end{tabular}
  \end{ruledtabular}
\end{table}

\section{results for all irreducible representations}
Each odd-parity representation can support topological crystalline superconducting states and associated Majorana surface states (MSSs)~\cite{Tei2023Possible}.
These states are protected by chiral symmetries constructed from twofold rotation or mirror reflection symmetries.
Consequently, the MSSs exhibit an Ising spin response whose direction is determined by the protecting crystalline symmetry.
Table~\ref{table:UTe2_topo} summarizes the MSSs, their protecting crystalline symmetries, and the corresponding Ising directions for all odd-parity IRs on the symmetry-preserving $(100)$, $(010)$, and $(001)$ surfaces.
The surface-dependent Majorana Ising direction can therefore serve as a powerful probe of the superconducting pairing symmetry.

\begin{figure}[p]
    \centering
    \includegraphics[width=\linewidth]{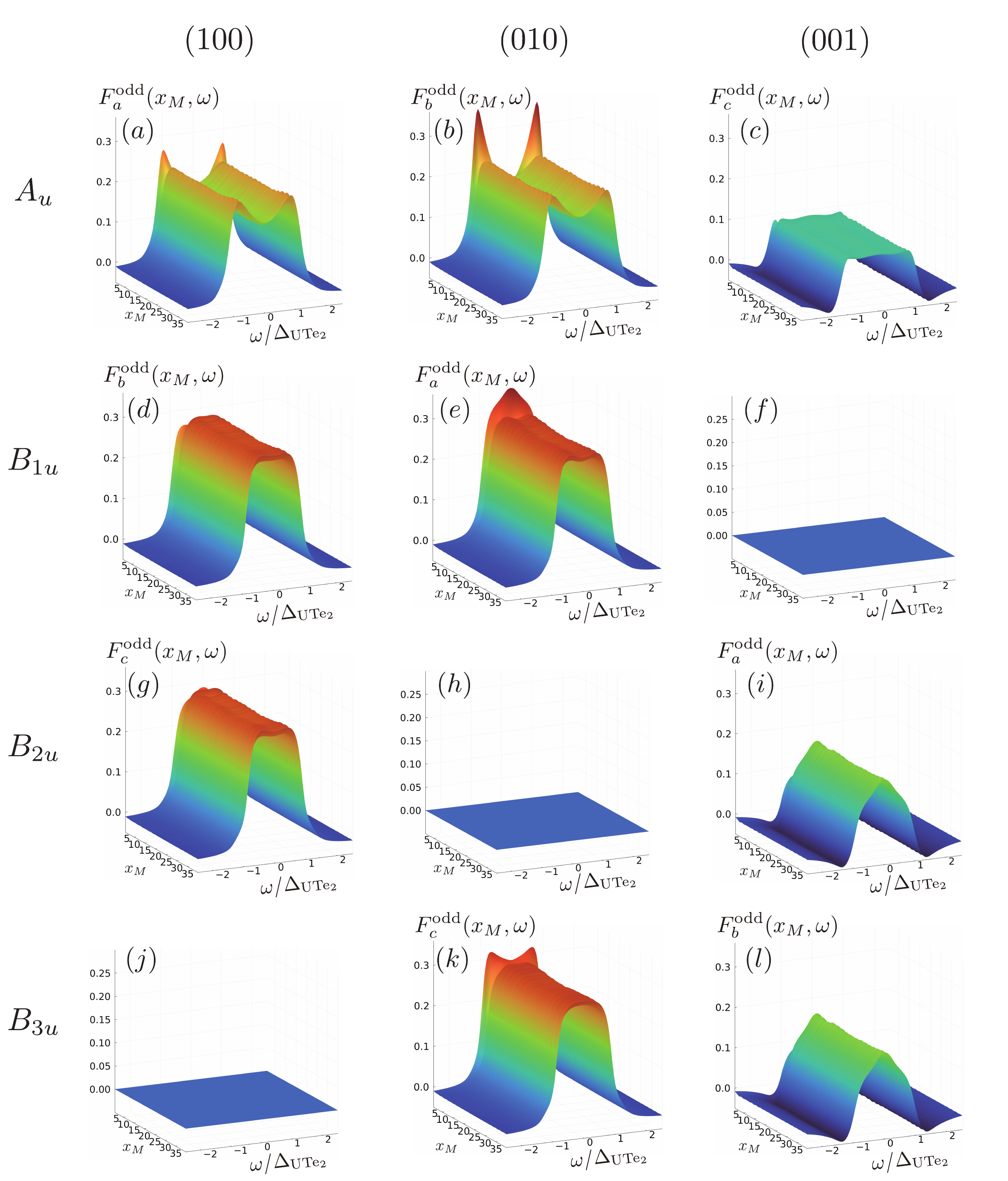}
    \caption{Spatial and frequency dependence of odd-frequency pairing proximitized in the metal.
    $x_M=1$ corresponds to the interface with the superconductor.
    (a)--(c), (d)--(f), (g)--(i), and (j)--(l) show the $A_u$, $B_{1u}$, $B_{2u}$, and $B_{3u}$ states, respectively.
    The first, second, and third columns show the $(100)$, $(010)$, and $(001)$ surfaces, respectively.}
    \label{figS1}
\end{figure}

Figure~\ref{figS1} summarizes the odd-frequency spin-triplet pairing amplitudes induced in the normal-metal film through the proximity effect.
We consider a diffusive metal film with $\ell_M/a=10$ and $N_M=40$.
We use a $64\times 64$ $\bm{k}_{\parallel}$-point mesh and an energy-smearing parameter of $\delta=0.01Z$.
For the $A_u$ state, topologically protected MSSs appear on the $(100)$ and $(010)$ surfaces, giving rise to pronounced odd-frequency pairing in the metal film.
The $(001)$ surface is topologically trivial and hosts no MSSs, although a weak odd-frequency component is still induced.
We note that odd-frequency pairing can generally arise from broken translational symmetry even in the absence of MSSs.
For the $B_{1u}$ state, MSSs on the $(100)$ and $(010)$ surfaces similarly give rise to pronounced odd-frequency pairing.
For the $B_{2u}$ state, MSSs are present on the $(100)$ and $(001)$ surfaces and induce odd-frequency pairing in the metal film.
However, on the $(001)$ surface, the projected area of the quasi-2D Fermi surfaces is small, leading to weak proximity coupling and consequently a small induced odd-frequency pairing amplitude.
Within the present model, the results for the $B_{3u}$ state are related to those for the $B_{2u}$ state by interchanging the $a$ and $b$ axes.

\begin{figure}[p]
    \centering
    \includegraphics[width=\linewidth]{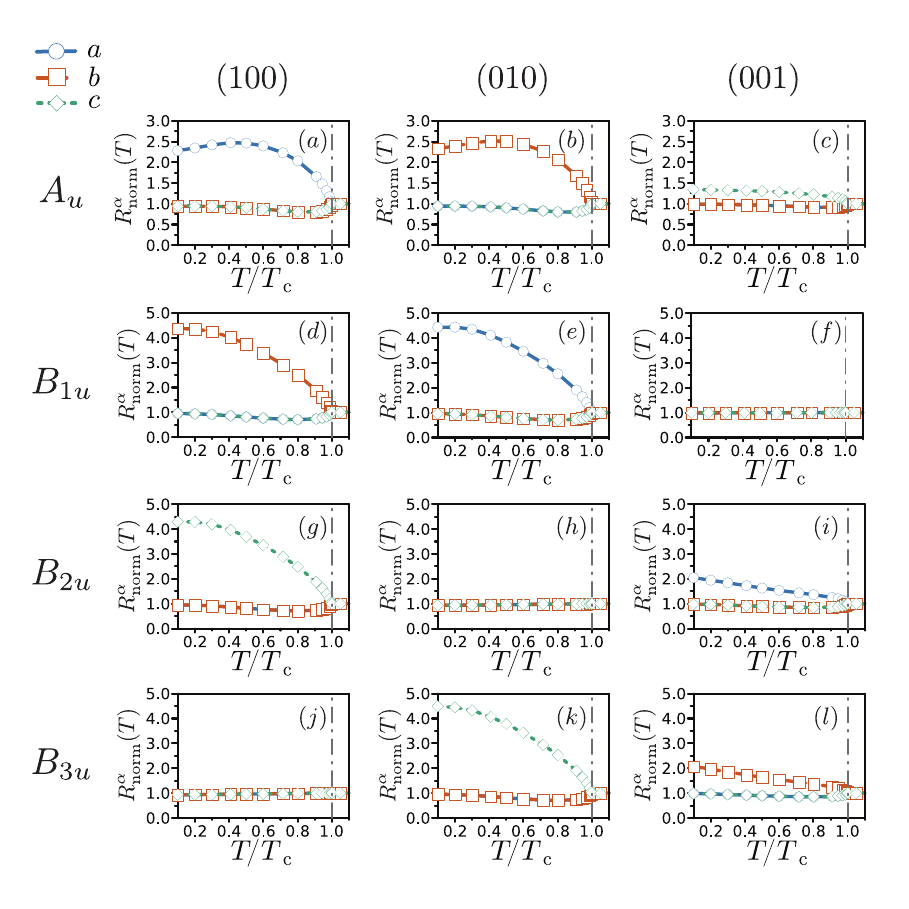}
    \caption{
    Temperature dependence of the normalized dynamical spin-response components 
    $R^{\alpha}_{\rm norm}(T)=R^{\alpha}(T)/R^{\alpha}(T_{\rm c})$.
    (a)--(c), (d)--(f), (g)--(i), and (j)--(l) show the $A_u$, $B_{1u}$, $B_{2u}$, and $B_{3u}$ states, respectively.
    The first, second, and third columns show the $(100)$, $(010)$, and $(001)$ surfaces, respectively.
    The metal film thickness is set to $N_M=40$ lattice layers.}
    \label{figS2}
\end{figure}

\begin{figure}[p]
    \centering
    \includegraphics[width=\linewidth]{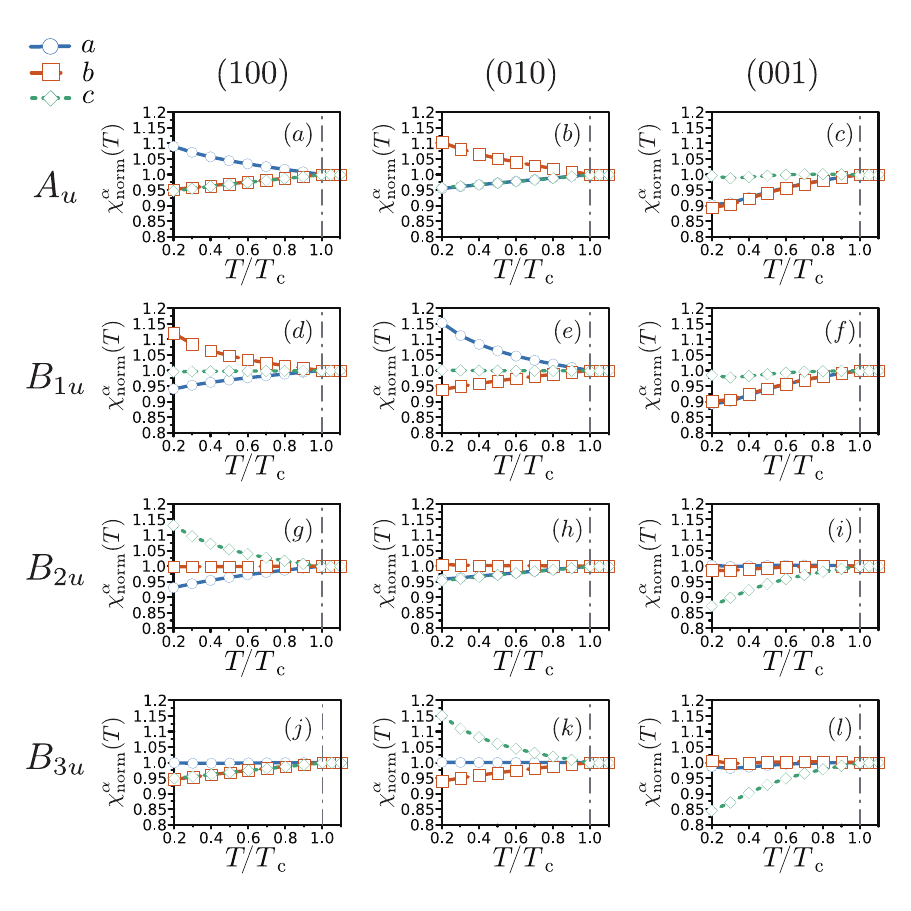}
    \caption{
    Temperature dependence of the normalized static spin-susceptibility components 
    $\chi^{\alpha}_{\rm norm}(T)=\chi^{\alpha}(T)/\chi^{\alpha}(T_{\rm c})$.
    (a)--(c), (d)--(f), (g)--(i), and (j)--(l) show the $A_u$, $B_{1u}$, $B_{2u}$, and $B_{3u}$ states, respectively.
    The first, second, and third columns show the $(100)$, $(010)$, and $(001)$ surfaces, respectively.
    The metal film thickness is set to $N_M=40$ lattice layers.}
    \label{figS3}
\end{figure}

Figures~\ref{figS2} and \ref{figS3} show the temperature dependence of the dynamical spin response $R^{\alpha}(T)$ and the static spin susceptibility $\chi^{\alpha}(T)$, respectively, for each IR and surface orientation.
$R^{\alpha}(T)$ is calculated in the diffusive regime with a short mean-free path, $\ell_M/a=10$, whereas $\chi^{\alpha}(T)$ is evaluated in the clean limit.
The metal film thickness is set to $N_M=40$ lattice layers.

\clearpage
\bibliography{paper}